\documentclass{article}
\usepackage[utf8]{inputenc}
\usepackage{authblk}
\usepackage{hyperref}
\usepackage{amsmath}
\usepackage{xcolor}

\usepackage[numbers]{natbib}
\usepackage{graphicx}

\newcommand\blfootnote[1]{%
  \begingroup
  \renewcommand\thefootnote{}\footnote{#1}%
  \addtocounter{footnote}{-1}%
  \endgroup
}

\title{Phase stability of dispersions of hollow silica nanocubes mediated by non-adsorbing polymers}
\author[1]{Frans Dekker} 
\author[1,2]{{\'A}lvaro Gonz{\'a}lez Garc{\'i}a} 
\author[1]{Albert P. Philipse}
\author[2]{Remco Tuinier$^{*}$}
\affil[1]{Van ’t Hoff Laboratory for Physical \& Colloid Chemistry, Department of Chemistry \& Debye Institute, Utrecht University, The Netherlands}
\affil[2]{Laboratory of Physical Chemistry, Department of Chemical Engineering \& Chemistry \& Institute for Complex Molecular Systems (ICMS), Eindhoven University of Technology, The Netherlands}

\begin{document}
\maketitle
\blfootnote{${^*}$ to whom correspondence should be addressed: r.tuinier@tue.nl}

\section*{Abstract}
Although there are theoretical predictions 
[\textit{Eur.~Phys.~J.~E} \textbf{41} (2018) 110] for the rich phase behaviour of colloidal cubes mixed with non-adsorbing polymers, a  thorough verification of this phase behaviour is still underway; experimental studies on mixtures of cubes and non-adsorbing polymers in bulk are scarce. In this paper, mixtures of hollow silica nanocubes and linear polystyrene in \textit{N,-N-}dimethylformamide are used to measure the structure factor of the colloidal cubes as a function of non-adsorbing polymer concentration. Together with visual observations these structure factors enabled us to assess the depletion-mediated phase stability of cube-polymer mixtures. The theoretical and experimental phase boundaries for cube-depletant mixtures are in remarkable agreement, despite the simplifications underlying the theory employed.

\vspace{5pt}
\hrule
\clearpage
\section{Introduction}
\label{intro}
Ordered structures prepared from functional colloids have multiple applications, and are expected to play an important role in the development of new technologies. Colloidal solids obtained from micron-sized particles exhibit a photonic bandgap \cite{Vlasov2001}, while plasmonic nanoparticle solids can be applied as highly sensitive sensors \cite{Anker2008}. There are different methods to prepare colloidal solids \cite{Lotito2017,Vogel2015}, but most of the available techniques require high particle concentrations or operate under out-of-equilibrium conditions \cite{Lotito2017,VanDerBurgt2018,Meijer2019}. To form assemblies with desired structural properties, a delicate control over the colloidal interactions is required.

Particle--particle interactions are affected by the addition of non-adsorbing polymers, often termed depletants \cite{Lekkerkerker2011}. Due to configurational entropy loss of the depleted polymer chains near the surfaces of the colloidal particles, the polymer segment density close to the colloidal particles is lower than in bulk \cite{Gonzalez2019}. The polymer concentration profile defines the depletion zones around colloidal particles in presence of non-adsorbing polymers \cite{Oosawa1954a,Asakura1958}. Whenever overlap of depletion zones occurs, there is an osmotic pressure difference between the bulk and the overlapping volume, leading to a net attraction between the colloidal particles. This depletion attraction between colloidal particles is, to a certain degree, tuneable via the depletant concentration, the size ratio between depletant and colloidal particle, and the shape of the colloidal particle \cite{Lekkerkerker2011,Vrij1976,Sacanna2010,Petukhov2017}.

The attraction between two bodies immersed in a solution of non-adsorbing macromolecules was first predicted theoretically by Asakura and Oosawa in the fifties \cite{Oosawa1954a}. Not much later, Sieglaff \cite{Sieglaf1959} demonstrated that addition of polystyrene to a dispersions of micro-gel spheres in toluene results in depletion-induced demixing. Systematic studies on the stability of particles in polymer solutions were first performed by Vincent \textit{et al.} \cite{Long1973,Lin-In-On1974,Cowell1978,Vincent1980}, who studied mixtures of latex spheres in aqueous polyethylene oxide (PEO) solutions. The direct link between experiments and theory on the phase stability was pioneered by de Hek and Vrij \cite{DeHek1981,DeHek1982}, who studied mixtures of silica spheres and polystyrene in cyclohexane.

Vrij \cite{Vrij1976} simplified polymers as penetrable hard spheres (PHSs): the interactions with the colloidal particles is hard-core, but they can freely interpenetrate each other. Based upon pair-wise additivity of the depletion interaction potential de Hek and Vrij \cite{DeHek1981} predicted the demixing concentrations of silica--polystyrene mixtures with varying molar mass $M$. Later, Gast \textit{et al.} \cite{Gast1983} developed a method to calculate phase diagrams of colloidal particles mixed with PHSs using thermodynamic perturbation theory from the pair-wise interactions. 
 
Lekkerkerker \cite{Lekkerkerker1990} proposed free volume theory (FVT), which corrects for multiple overlap of depletion layers. The phase diagrams calculated using FVT shows that demixing of a single phase in colloid--polymer mixtures takes place. The predicted possible phase coexistences include colloidal fluid--fluid (or gas--liquid) as well as fluid--solid equilibria. Remarkably, a three-phase  colloidal gas--liquid--solid coexistence \cite{Lekkerkerker1992} is also predicted, which was experimentally demonstrated in colloid-polymer mixtures \cite{Ilett1995,Faers1997}.

Attention has been given to mixtures of anisotropic colloidal particles and depletants. The depletion interaction and resulting phase separation of rod-like particles has been used to concentrate viruses, such as tobacco mosaic virus (TMV) \cite{Venekamp1964}. FVT predictions on rod-depletant mixtures \cite{Lekkerkerker1994}  were used to quantify the phase behaviour of grafted boehmite rods with polystyrene polymers in dichlorobenzene  \cite{Buitenhuis1995}, as well as aqueous mixtures of TMV viruses and polyethylene oxide \cite{Adams1998}. The predictions were confirmed using computer simulations \cite{Bolhuis1997,Savenko2006}. 

Plate-like particles are also often studied since they are ubiquitous in nature \cite{Lekkerkerker2013,Liu2017}. Van der Kooij \textit{et al.} \cite{VanderKooij2000} demonstrated that addition of non-adsorbing polymers to gibbsite platelets resulted in the formation of a dense columnar phase  and even lead to a four phase-coexistence \cite{VanderKooij2000}. This rich phase behaviour was later also observed in Mg:Al oxide platelets \cite{Zhu2007}, and tackled from simulations and theory \cite{Zhang2002,GonzalezGarcia2018b}. Colloidal plates mixed with colloidal spheres also manifest depletion-induced phase separation \cite{Kleshchanok2010,Doshi2011}.

\begin{figure}
\resizebox{0.95\linewidth}{!}{%
  \includegraphics{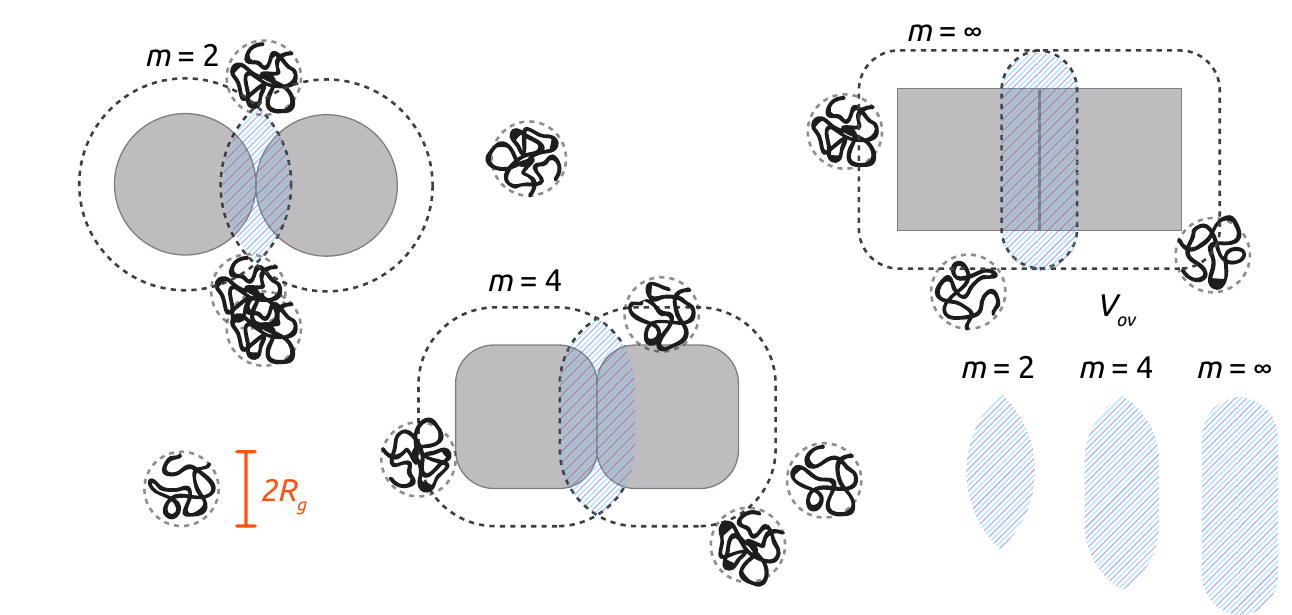}}
\caption{Sketches of different types of two hard superballs in face-to-face contact in non-adsorbing polymer solutions. The presence of the polymers leads to depletion zones (dashed layers around the particles). The hatched areas reflect overlap volumes of depletion zones. The examples are given for superballs with shape parameters $m = 2$, 4, and $\infty$.}
\label{fig:1} 
\end{figure}

Previous experiments on dispersions of micron-sized superballs with added non-adsorbing polymers revealed a rich phase behaviour. It was demonstrated that the obtained phase states of the sediment depend on the colloid--depletant size ratio \cite{Rossi2011} and the details of the shape of the cube-like colloidal particles. \cite{Rossi2015}. Although the phase diagram of cubes and polymers was recently predicted theoretically \cite{GonzalezGarcia2018}, experimental studies on the phase behaviour of \textit{stable} dispersions of colloidal nanocubes mixed with non-adsorbing polymers are still lacking. Recently, we showed that hollow silica nanocubes display effective hard-core interactions, which makes them promising particles to study the effect of non-adsorbing polymers on the phase behaviour of model anisotropic particles \cite{Dekker2019b}.The nanocube-like particles were mapped onto the so-called superball shape, whose surface is defined by the locus of points that satisfy
\begin{align}
    |2x/R_\text{el}|^m+|2y/R_\text{el}|^m+|2z/R_\text{el}|^m = 1 \text{ ,}
\end{align}
where $R_\text{el}$ is the edge length of the superball and $m$ is the shape parameter \cite{Dekker2019b}. For $m=2$ a sphere ($R_\mathrm{el} = 2R_\mathrm{sphere}$) is recovered and $m=\infty$ corresponds to a cube. In fig. \ref{fig:1},
the overlap volume for superballs, with shape parameters $m = 2, 4$, and $\infty$ is depicted schematically. The depletion overlap volume increases with  $m$ when superballs align “face to face”: depletant addition enhances the well-known tendency of flat faces to align \cite{Petukhov2017}. 

In this paper, we assess the depletion-induced phase separation on the otherwise stable silica nanocube dispersions using static light scattering (SLS). First, we outline the preparation of the cubic nanoparticles. Then, we discuss how SLS can be used to monitor the stability of colloid--polymer mixtures. We present an experimental method to obtain stable cube fluids and scattering curves as a function of the depletant concentration. Subsequently, a brief explanation is provided about how free volume theory (FVT) is used to predict the phase stability of cube-polymer mixtures. We conclude testing the theoretical predictions against the first experimental phase diagram of a stable nanocube fluid mixed with non-adsorbing polymers.

\section{Materials and methods}
\subsection{Cubic silica shell in DMF}
\label{sec:8}
\textit{N,-N-}Dimethylformamide (DMF, Anhydrous 99.8\%) was purchased from Sigma-Aldrich and LiCl (Anhydrous, 99\%) was acquired from Alfa Aesar. Polystyrene ($M_\text{w} = 600$ $\text{kg/mol}$, $M_\text{w}/M_\text{n} < 1.10$) was obtained from Pressure Chemical. Stable dispersions of hollow silica nanocubes were prepared as described in \cite{Dekker2018}. In short, cuprous oxide nanocubes were prepared using the polyol method \cite{Park2009} and subsequently coated with St{\"o}ber silica in the presence of PVP \cite{Graf2003,Castillo2014}. The cuprous oxide core was removed with a mixture of nitric acid and hydrochloric acid and the particles were dispersed in DMF by repeated washing using centrifugation steps. In a final step the particles were dispersed in DMF containing 40 mM LiCl to set the Debye screening length to $\kappa^{-1} \approx 1$ nm. The stability and specific volume of the cubic silica shells were determined by static light scattering as described in \cite{Dekker2019b}.

\subsection{Experimental assessment of phase stability}
\label{sec:5}
Experimental determination of phase stability is often done via visual inspection or using light microscopy \cite{Rossi2015,VanderKooij2000,Zhou2010}. Additionally, scattering methods such as static light scattering (SLS) \cite{Bodnar1996,DeHek1982}, small-angle neutron scattering (SANS) \cite{Kumar2013,Ye1996,Ye1996a} or small-angle X-ray scattering (SAXS) \cite{VanderKooij2000} can be employed to indirectly measure interactions between colloidal particles, to determine the phase stability, or to characterise the formed phases.  \textcolor{black}{Employing a scattering method such as SLS in addition to visual observation allows to determine the onset of phase separation. In principle, scattering experiments can be used to estimate the spinodal \cite{DeHek1981} and quantify the structure of the dispersion, and hence also enable to determine the second virial coefficient $B_2$ of colloidal particles in the presence of depletants \cite{DeHek1982}.} In the Rayleigh-Gans-Debye approximation the normalised scattering intensity $R(K)$ of a single component monodisperse spherical particles in a background solvent is given by \cite{Kerker1997}:
\begin{equation}
R(K) = \rho_{i} H_i P_i(K)S_i(K).
\label{eq:6}
\end{equation}
Here $P(K)$ is the particle form factor, $S(K)$ is the static structure factor, $\rho_i$ is the number density of component $i$. For a given solvent the contrast factor $H_i$ is given by:
\begin{equation}
H_i = \frac{2\mathrm{\pi}^2 v_{i}^2(n_\text{i}-n_\text{s})^2  n_\text{m}^2}{\lambda_0^4} \text{ ,}
\label{eq:7}
\end{equation}
where $v_i$, and $n_i$ respectively are the volume and refractive index of component $i$, $n_\text{s}$ the refractive index of the solvent, $n_\text{m}$ the refractive index of the mixture and $\lambda_0$ the wavelength of the incident light in vacuum.
For two-component dispersions in a background solvent composed of, for instance colloidal particles (c) and depletants (d), eq. \ref{eq:7} can be extended to \cite{Rodriguez1992}:
\begin{multline}
  R(K) = \rho_\text{c} H_\text{c} P_\text{c}(K)S_\text{cc}(K) + \rho_\text{d} H_\text{d} P_\text{d}(K)S_\text{dd}(K)+ \\
 2(H_\text{c}H_\text{d})^{1/2} (\rho_\text{c}\rho_\text{d})^{1/2} (P_\text{c}(K)P_\text{d}(K))^{1/2} S_\text{cd}(K).
\label{eq:8}
\end{multline}
In eq. \ref{eq:8}, $S_\text{cc}(K)$, $S_\text{dd}(K)$, and $S_\text{cd}(K)$ are the scattering contribution from the colloid--colloid, depletant--depletant and colloid--depletant interactions, respectively. With SANS the depletant and solvent can be chosen such that they have the same scattering length densities, making the depletant `invisible' for the neutrons ($H_\text{d} \rightarrow 0$), reducing eq. \ref{eq:8} to eq. \ref{eq:6} for a single component \cite{Ye1996a}. For SLS, rendering the depletant invisible could be achieved by index matching the solvent to the polymer. Index matching of polystyrene is difficult, however, because of the high refractive index of polystyrene. Although theory is available for the scattering of multicomponent systems \cite{Vrij2000} it is not straightforward to extend this to the scattering of cubic particles. SANS data on colloids with “invisible” depletants exhibit a significant increase in the $S(K)$ of the colloids for low $K$ values 
($KR < 1$), indicating attraction between the colloidal particles \cite{Vrij2005}. Since the hollow silica cubes are expected to scatter significantly more than the polymer chains in the low K limit, the scattering is expected to be dominated by the colloids. To extract an “apparent structure factor” $S^*(K)$, a similar route is taken as discussed in \cite{Dekker2019b}, using
\begin{equation}
S^*(K) = \frac{R(K,c_\text{c},c_\text{d})}{R(K,c_{\text{c},0})}\frac{c_{\text{c},0}}{c_\text{c}}.
\label{eq:9}
\end{equation}
Here $R(K,c_\text{c},c_\text{d})$ is the Rayleigh ratio for a particle concentration $c_\text{c}$ and depletant concentration $c_\text{d}$, and $R(K,c_c,0)$ is the Rayleigh ratio in the dilute limit (where $c_\text{c}\rightarrow c_\text{c,0}$) at sufficiently low $c_\text{c}$ $(cc = c_\text{c,0})$, where $S(K) \equiv 1$. The apparent structure factor can then be obtained as a function of the depletant concentrations to determine the influence on the stability of the hollow silica cubes.

The onset of phase separation was monitored with SLS. Typically, the following procedure was employed. To a clean and dust-free cuvette, a weighed amount of DMF (with 40 mM LiCl) was added containing 1.8 g/L silica nanocubes. The scattering curve of this dispersion was measured to obtain the experimental form factor. The concentration of this dispersion was then increased to 24 g/L by addition of a concentrated stock dispersion in the same solvent, of which the scattering curve was also obtained to obtain the structure factor at finite concentration. To the 0.5 mL of 24 g/L silica cube dispersion, 30 $\mu L$ polystyrene (39 g/L in DMF containing 40 mM LiCl) was added, after which the scattering intensity was measured. This was repeated until phase separation was observed visually and could also be derived from the scattering curves. The dispersion was then diluted with DMF (0.2 mL, 40 mM) to obtain a stable dispersion again, of which also the scattering curve was obtained. \textcolor{black}{In Table \ref{tab:1} an overview is presented of the four different depletion experiments (DEP 1 -- DEP 4) that were conducted. The concentrations of HSN and PS of the studied mixtures are listed in Table \ref{tab:1} in g/L.} Reversibility of phase separation upon diluting and concentrating a colloid-polymer mixture is characteristic of depletion-induced demixing \cite{Lekkerkerker2011}.

\subsection{Depletion attraction: linking theory and experiments}
\label{sec:2}
Simplifying the polymers as penetrable hard spheres (PHSs)  provides an analytic expression for the interaction potential between two hard colloidal spheres with radius $a$ when dispersed in a solution containing non-adsorbing macromolecules with radius of gyration $R_g$ \cite{Lekkerkerker2011}.
\begin{align*}
&&  		W_\text{s}(r) 	&= \infty 							&  		&r < 2a & &\\
&&					&= -\mathit{\Pi}_\text{d} V_\text{ov}(r) 	&   2a \leq \;  &r \leq 2R_\text{d}  & & \\
&& 					&= 0 								& 		&r > 2R_\text{d},  & & 
\label{eq:1}
\end{align*}
with the overlap volume:
\begin{equation}
V_\text{ov}(r)=\frac{4\mathrm{\pi}}{3}R_\text{d}^3 \left(1-\frac{3 r}{4R_\text{d}} + \frac{ r^3}{16 R_\text{d}^3} \right),
\label{eq:2}
\end{equation}
where $r$ is the centre-to-centre distance between the two colloidal hard spheres and interaction radius $R_\text{d} = a + R_\text{g}$.Here, it is assumed that the depletion thickness is such that $\delta=R_\text{g}$.

In the dilute limit the osmotic pressure of the depletants is given by Van ‘t Hoff's osmotic pressure law:
\begin{equation}
\mathit{\Pi}_\text{d}=\rho_\text{d} k_\text{B}T = \frac{\phi_\text{d}}{v_\text{d}} k_\text{B}T,
\label{eq:3}
\end{equation}
where $\rho_\text{d}$ is the number density of the depletants, $k_\mathrm{B}$ the Boltzmann’s constant, 
$T$ the absolute temperature and $v_\text{d}$ the volume of the depletant, here taken as $4 \pi R_\text{g}^3/3$. The relative depletant concentration $\phi_\text{d}$ is often expressed in terms of the polymer concentration $C_\text{d}$ and the depletant overlap concentration $C*$:
\begin{equation}
\phi_\text{d} = \frac{C_\text{d}}{C^*}.
\label{eq:4}
\end{equation}
The polymer overlap concentration can be estimated using
\begin{equation}
C^* = \frac{M}{v_\text{d}N_\text{A}} = \frac{3 M}{4\mathrm{\pi} R_\text{g}^3 N_\text{A}},
\label{eq:5}
\end{equation}
where $M$ is the molar mass of the polymer and $N_{A}$ is Avogadro’s constant. Eq. \ref{eq:3} shows that $\Pi_\text{d}$, which determines the strength of the depletion interaction, scales linearly with $\phi_\text{d}$, while the range of the interaction is determined by the value of $R_g$. 

\subsection{Free volume theory for nanocube--depletant mixtures}
\label{sec:3}
\begin{figure}
\resizebox{0.95\linewidth}{!}{%
  \includegraphics{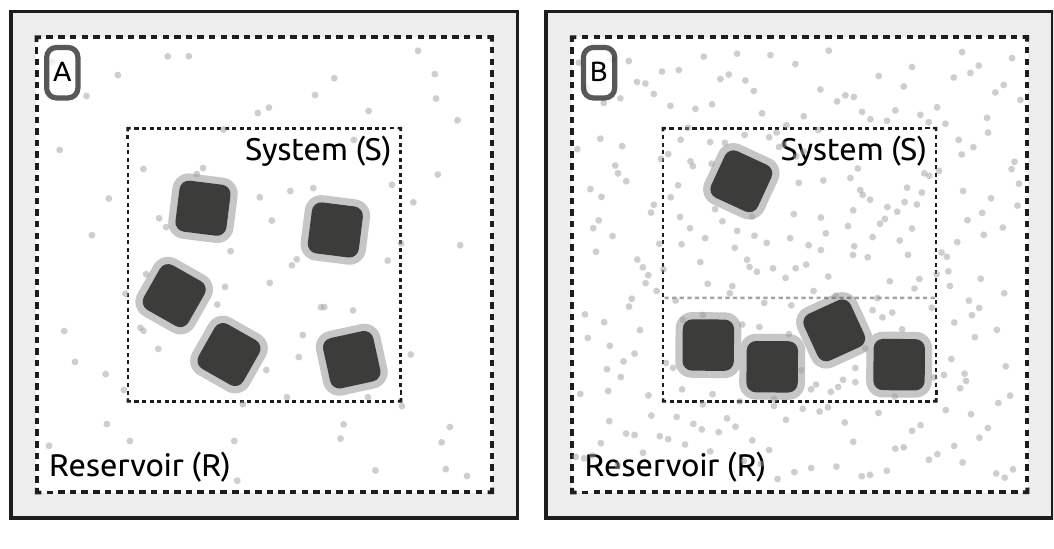}}
\caption{Schematic representation of the system (S) and reservoir (R) used in free volume theory (FVT). A:  single-phase system (low depletant concentration), B: FVT phase-separated system (high amount of depletants), with partitioning of depletants and colloidal particles over the phases.}
\label{fig:2} 
\end{figure}

Free Volume Theory (FVT) \cite{Lekkerkerker1992} allows constructing the equilibrium phase behaviour of colloid--depletant mixtures. Here we present a brief qualitative overview of the concepts behind FVT. For an in-depth discussion of the theory, see Ref. \cite{Lekkerkerker1992}. Depletants are described as penetrable hard spheres (PHSs) and colloids as hard superballs. FVT describes a system $(S)$ consisting of $N_\text{p}$ colloidal particles and $N_\text{d}$ depletants in a background solvent in equilibrium with a reservoir $(R)$ which does not contain colloidal particles, see the sketch in fig. \ref{fig:2}. System and reservoir are separated by a semi-permeable membrane that allows free exchange of solvent and depletants, but which is impermeable for the colloidal particles. The FVT construction leads to a semi-grand potential, which comprises the chemical potential and osmotic pressure of the colloidal particles in the different possible particle phase states. With these expressions it is possible to calculate phase coexistence curves and construct phase diagrams depending on the relative depletant volume fraction $\phi_\text{d}$, colloid volume fraction $\phi_\text{c}$, depletant-to-colloid size ratio $q$ and shape parameter $m$. To link experimental results to theoretical predictions, it is required to estimate these parameters. Determining $\phi_\text{c}$, $R$ and $m$ for cubic silica shells was done according to \cite{Dekker2018} and \cite{Dekker2019}. By determining the radius of gyration $R_\text{g}$ of the polymer chains in solution with light scattering, the size ratio $q$ can be estimated. 

The normalised semi-grand potential of the system ($\tilde{\mathit{\Omega}}$) can be written as \cite{Lekkerkerker2011}:
\begin{equation}
\tilde{\mathit{\Omega}} = \frac{\mathit{\Omega} v_\text{c}}{k_\text{B}TV} =\tilde{A}_0-\tilde{\mathit{\Pi}}_\text{d}^\text{R} \frac{v_\text{c}}{v_\text{d}} \alpha.
\label{eq:10}
\end{equation}
Here, $v_\text{c}$ and $v_\text{d}$ are the particle and depletant volumes, $V$ is the volume of the system, $A_0$ is the Helmholtz energy of the pure particles in the system, $\tilde{A}_0 = A_0 v_\text{d}/(k_\mathrm{B}TV)$, $\Pi_\text{d}^\text{R}$ the osmotic pressure of the depletants in the reservoir and $\alpha$ the free volume fraction for the depletants in the system. Since the polymers are considered to behave ideally, the osmotic pressure is given by Van ’t Hoff expression:
\begin{equation}
\tilde{\mathit{\Pi}}_\text{d}^\text{R} = \frac{\mathit{\Pi}_\text{d}^\text{R} v_\text{d}}{k_\text{B}T} = \phi_\text{d},
\label{eq:11}
\end{equation}
where $\phi_\text{d}$ is the relative concentration of depletants in the reservoir and $v_\text{d}$ the volume of the depletant. The free energy of the particles $A_0$ can be calculated for many hard particle fluids and some particle solids. For superballs, expressions for the free energy of particles in the fluid, face-centered cubic, and simple cubic phases were proposed previously \cite{GonzalezGarcia2018}. Further we need to quantify $\alpha$ in eq. \ref{eq:10}. The free volume fraction $\alpha$ is formally given by the reversible work $w$ required to put a depletant from the reservoir into the system as
\begin{equation}
\alpha = \mathrm{e}^{-\mathit{w}/k_\text{B}T}.
\label{eq:12}
\end{equation}
This required work can be estimated using scaled particle theory \cite{Lekkerkerker2011}. With this expression for $\alpha$ it is possible to calculate the chemical potential $\mu_\text{c}$ of the colloidal particles and the osmotic pressure in the colloid--polymer mixture via
\begin{equation}
\tilde{\mu}_\text{c} = \frac{\mu_\text{c}}{k_\text{B}T} = \left( \frac{\partial\tilde{\mathit{\Omega}}}{\partial\phi_\text{c}} \right)_{T,V,N_\text{d}^\text{R}}  \quad ; \quad \tilde{\mathit{\Pi}} = \frac{\mathit{\Pi} v_\text{c}}{k_\text{B}T} = \phi_\text{c} \tilde{\mu} - \tilde{\mathit{\Omega}},
\label{eq:13}
\end{equation}
where $N_\text{d}^\text{R}$ the number of depletants in $R$.  At a given depletant concentration it can be verified whether phase coexistence takes place by determining the osmotic pressure and chemical potentials of the phases considered are equal. For further details, see \cite{GonzalezGarcia2018}.

\clearpage
\section{Results}
\label{sec:11}
\subsection{Size ratio}
\label{sec:12}
\begin{figure*}
\centering
  \includegraphics[width=0.8\linewidth]{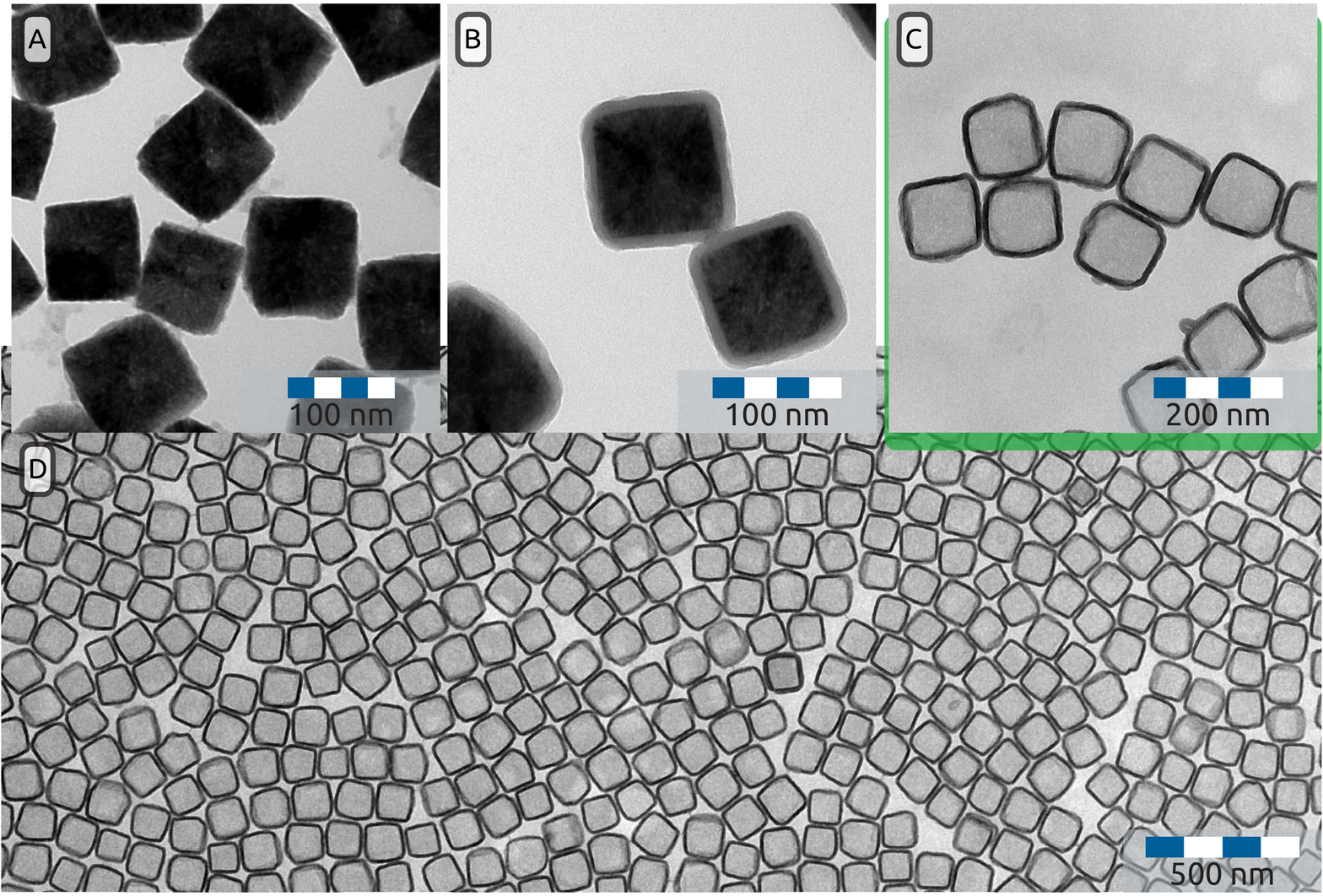}
\caption{TEM micrographs of the nanocubes that were used for the scattering and phase stability experiments. A: Cu\textsubscript{2}O nanocubes with an average size $R_\text{el} = 105 \pm 10$ nm. B: SiO\textsubscript{2}@Cu\textsubscript{2}O core-shell nanocubes with an average size $R_\text{el} = 125 \pm 10$ nm. C: Hollow SiO\textsubscript{2} nanocubes with an average size $R_\text{el} = 125 \pm 10$ nm and an $m$ value of $4.1 \pm 0.6$. D: Overview micrograph of many similar particles as those depicted in C.}
\label{fig:3} 
\end{figure*}

\begin{figure}
\resizebox{0.75\linewidth}{!}{%
  \includegraphics{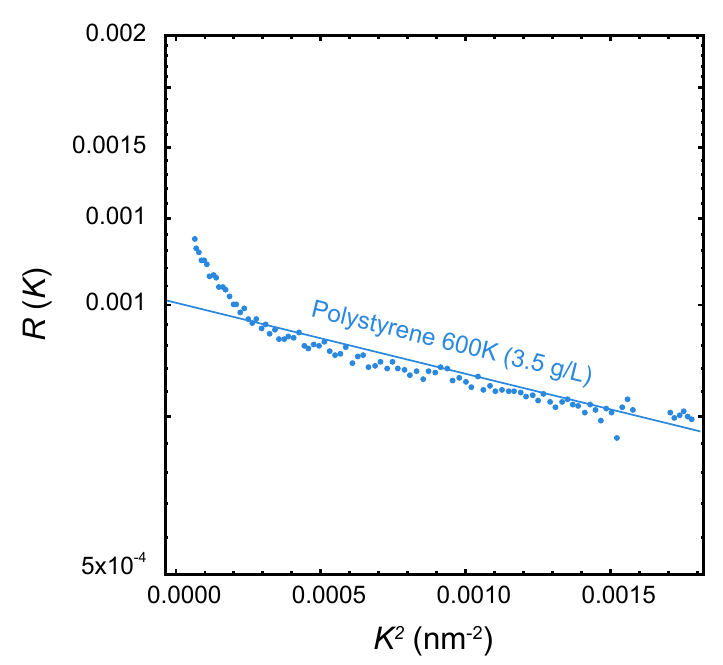}}
\caption{Guinier plot of the polystyrene used for the phase separation measurements. The blue line is a linear fit to determine the radius of gyration.}
\label{fig:4} 
\end{figure}

The size of the hollow silica nanocubes (HSN) was determined with transmission electron microscopy; representative micrographs are depicted in fig. \ref{fig:3}. The cubes have an average edge length $R_\text{el} = 125 \pm 10$ nm with a cubicity shape parameter $m$ of $4.1 \pm 0.6$. In order to determine the size ratio $q$, the radius of gyration of the polystyrene in DMF was determined with SLS. In fig. \ref{fig:4} a Guinier plot of polystyrene with $M_\text{w} = 600$ $\text{kg/mol}$ is depicted, from which we obtain $R_\text{g} = 21.5 \pm 1.1$ nm, resulting in $q = 2R_\text{g} / R_\text{el} = 0.34$. \textcolor{black}{The increased scattering at low scattering vectors originates from impurities, such as dust particles, present in the polystyrene stock solution. These impurities are difficult to remove from the stock solution, and since the scattering of the particles at these $K$ vectors is significantly higher, we surmise the scattering of the impurities do not significantly influence the results.}

\subsection{Phase transition from SLS}
\label{sec:13}
\begin{figure*}[tb!]
  \includegraphics[width=0.8\linewidth]{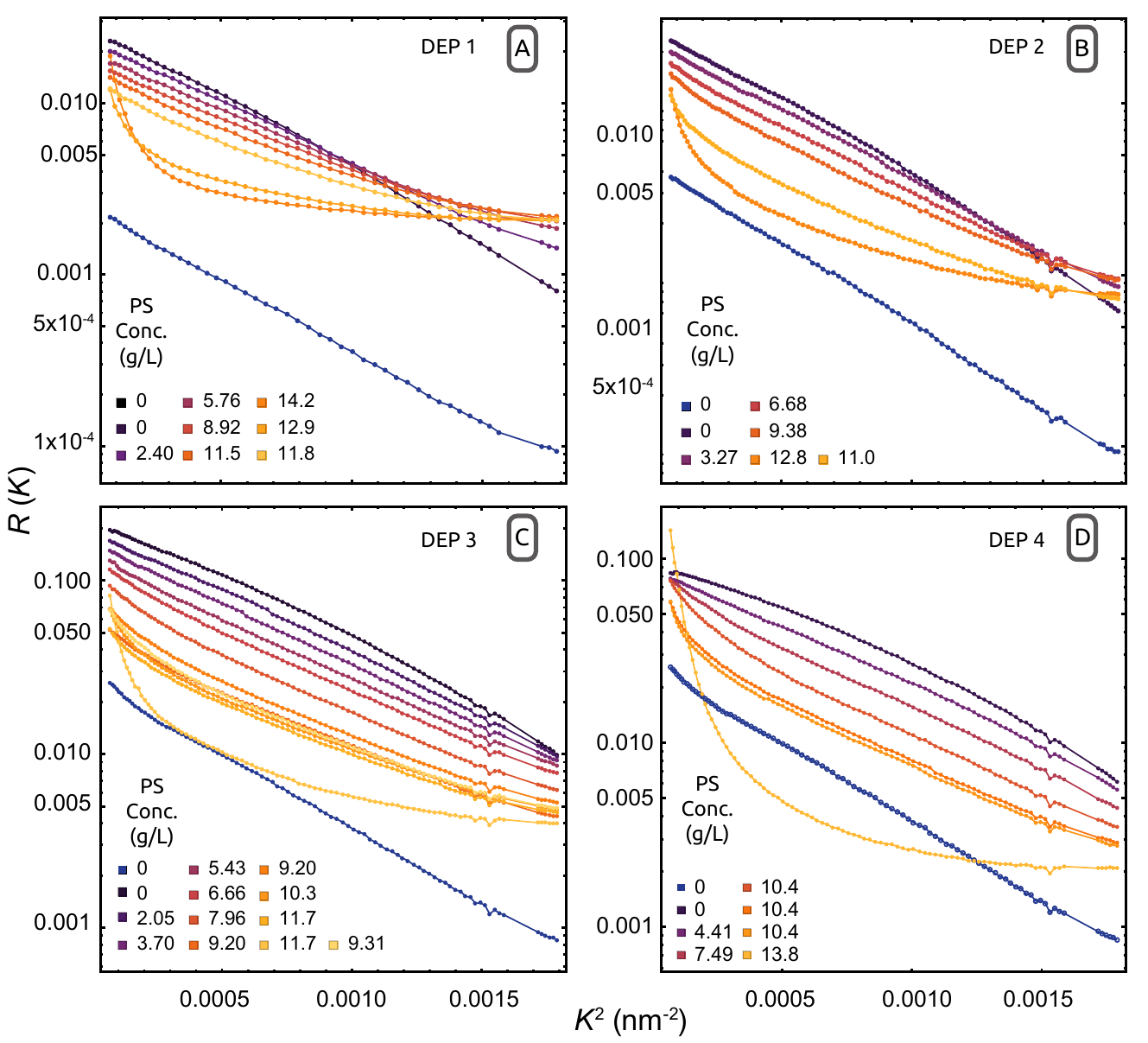}
  \centering
\caption{Scattered intensity $R(K)$ plotted against the scattering wave vector $K$ for various concentrations of HSN and various polystyrene concentrations, as indicated within the plots. The lines are to guide the eye. A: $R(K)$ for nanocube concentrations starting at $4.6$ g/L. B: $R(K)$ for nanocube concentrations starting at $10$ g/L. C: $R(K)$ for nanocube concentrations starting at $24$ g/L. D: $R(K)$ for nanocube concentrations starting at $37$ g/L. \textcolor{black}{Graphs C and D contain measurements of dispersions containing a single concentration that was measured multiple times, (11.7 g/L in C and 10.4 g/L in D). These extra measurements were preformed because visual observation showed an increase in sample turbidity.}}
\label{fig:5}      
\end{figure*}

\begin{figure*}[tb!]
  \includegraphics[width=0.8\linewidth]{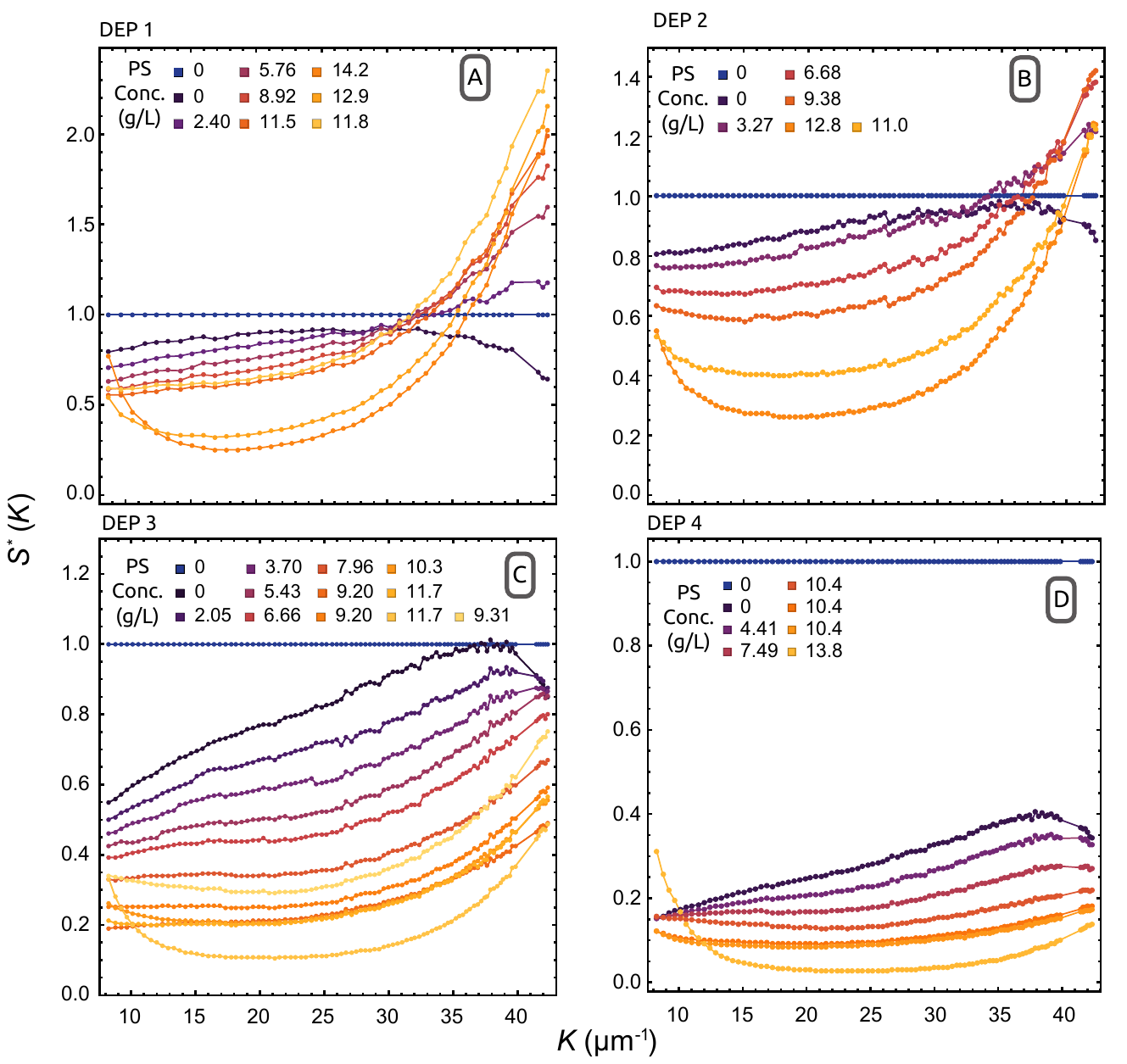}
  \centering
\caption{Apparent \textcolor{black}{structure factors $S^*(K)$ plotted} as a function of scattering wave vector $K$ for various concentrations of HSN and polystyrene. The curves are to guide the eye. A-D: \textcolor{black}{as in fig.\ref{fig:5}.}}
\label{fig:6}       
\end{figure*}

To determine the concentration at which the nanocubes and polymers demix, scattering curves of mixtures of nanocubes with various concentrations of added polymers were measured with SLS. Guinier plots obtained from these experiments are depicted in Figure \ref{fig:5}. At zero polymer concentration, a typical scattering curve for hollow silica particles is obtained, which is comparable to the scattering curves we reported earlier \cite{Dekker2019b}. When the polymer concentration is increased, initially two trends can be identified based on the scattering curves. The first is an overall decrease of the scattering intensity, which we assume is a result of polymer–polymer [$S(K)_\text{dd}$] and polymer–particle [$S(K)_\text{cd}$] interactions. Second, the scattering curves at high scattering vectors shift upward upon increasing the polystyrene concentration. This upward shift is especially visible for the experiments at low particle concentration (fig. \ref{fig:5}-A and fig. \ref{fig:5}-B). This increased scattered intensity is caused by the presence of polymers, which scatter more light at high $K$ compared to the hollow silica particles (fig. \ref{fig:s1}). 
At increasing polymer concentrations, the scattering curves bend upward for low $K$-values, indicating increasing mediated attractions between the nanocubes. Further addition of polymer leads to a further increase in scattering at low $K$ values with a corresponding drop in intensity at intermediate values for $K$. At a certain concentration the low $K$ data seem to diverge at $K \rightarrow 0$ indicating the mixture gets unstable since at the spinodal, $S (K\rightarrow 0)$ diverges.
To estimate at which concentrations demixing occurs, the apparent structure factor $S^*(K)$ was obtained by dividing the scattering curves with the scattering curve of hollow silica cubes at lowest concentration, and correcting for the concentration (eq. \ref{eq:9}). In fig. \ref{fig:6}, similar trends are visible in $S^*(K)$ compared to the discussed Guinier plots of fig. \ref{fig:5} the reduction in scattered intensity and the curves shifting upwards at large scattering vectors. The tendency to demix is clearly visible by a drop in $S^*(K)$ over a wide $K$ range and an upswing at low $K$-values towards an apparent divergence at $K \rightarrow 0$.
The phase separation was also visually observed. Furthermore, when the phase-separated mixture is diluted with DMF ($40$ mM LiCl), a stable dispersion is recovered, as follows from the apparent structure factors, and could also be concluded from visual observation. 

\subsection{Experimental Phase Diagrams}
\label{sec:14}
\begin{figure*}
  \includegraphics[width=0.8\linewidth]{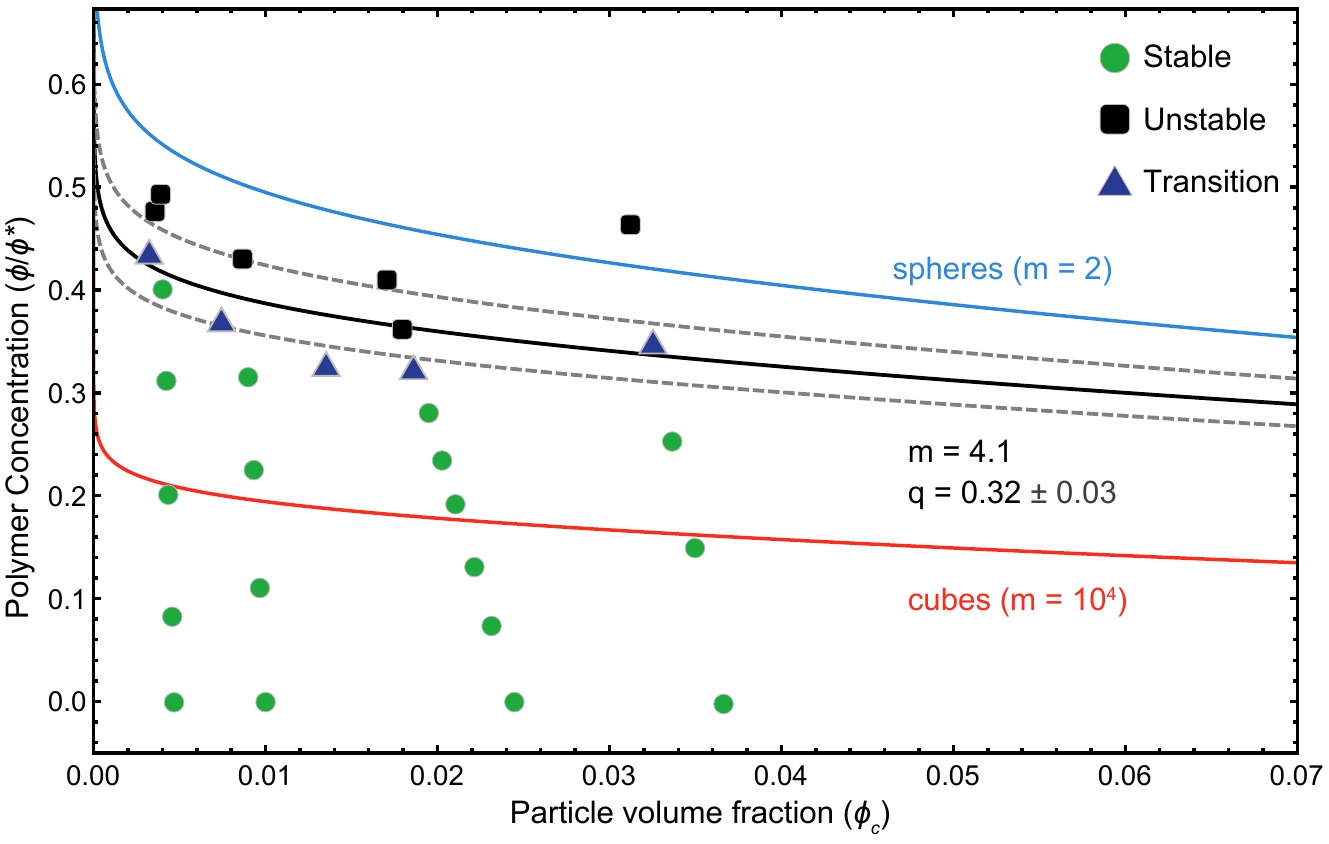}
  \centering
\caption{Experimental phase diagram of HSN and polystyrene in DMF with $40$ mM LiCl compared to theoretical predictions (curves) of hard superballs with non-adsorbing polymers. \textcolor{black}{The points follow from the raw data listed in Table \ref{tab:1}.} The curves are predicted fluid-solid coexistence binodal curves for superballs with $m$-values equal to $2$ (blue), $4.1$ (black), and $10^4$ (red) in solution containing non adsorbing polymers with size ratio $q = 2R_g / R_\text{el} = 0.32$. The grey dashed curves are binodal curves for superballs with $m = 4.1$ and size ratios $q = 2R_g/R_\text{el} = 0.29$ (bottom) and $q = 2R_g/R_\text{el}$ =0.35 (top).}
\label{fig:7} 
\end{figure*}

Based on the estimated phase-transition points from SLS we construct an experimental phase diagram for mixtures of HSN and polystyrene polymers \textcolor{black}{(PS)} in DMF ($40$ mM LiCl). \textcolor{black}{The dilution lines are constructed from the HSN and PS concentrations as listed in Table \ref{tab:1}.} In the phase diagram, depicted in fig. \ref{fig:7} we discern three different regions: a concentration range where the mixture is stable (depicted by green data points in fig. \ref{fig:7}), a concentration range where the mixture clearly phase separates (depicted by the black data points), and an intermediate transition region where no clear phase separation occurs, but where the $S^*(K)$ curves indicate that significant attraction is present (blue data points). In fig. \ref{fig:7} these regions are indicated, resulting in the first experimental phase diagram of colloidal nanocube dispersions with depletants we are aware of. 

As discussed previously, the access to model dispersions of cubes and polymers, where the interaction can be controlled by the depletion interaction, allows us to prepare colloidal solids of different phase states and compositions, depending on the exact shape and size of the colloids. The parameter space of the full phase diagram of cubes and polymers can be described by four key parameters; $\phi_\text{c}$, $\phi_\text{d}$, $m$ and $q$. This complete set of parameters is almost impossible to fully explore experimentally. Therefore, the availability of a theoretical framework that successfully predicts the phase behaviour of colloids and depletants is paramount for making progress in technical applications. 

In fig. \ref{fig:7} the experimental data are compared to theoretical predictions. Particle volume fractions were calculated using the specific volumes obtained earlier \cite{Dekker2019b} and the overlap concentration of polystyrene ($M_\text{w} = 600$ $\text{kg/mol}$, $R_\text{g} = 21.5 \pm 1.1$ nm) was calculated from eq. \ref{eq:5} and determined to be 29.7 g/L. The curves correspond to the fluid branch of the fluid-solid coexistence binodal for superballs with $m = 2$ (blue curve), $m=4.1$ (black curve) and $m=10^4$ (red curve) plus added non-adsorbing polymers with size ratio $q = 0.32$. The grey dashed curves are coexistence lines for superballs with $m = 4.1$ and size ratios $q = 0.29$ and $0.35$, representing the lower and upper limit of the polymer polydispersity. The experimental data are in remarkable agreement with the theoretical predictions, indicating that the theory is able to predict the depletion effects in experimental model systems and that dispersions of hollow silica cubes in DMF with 40 mM LiCl and polystyrene is such a model system.

\clearpage
\section{Conclusions}
\label{sec:15}
We have studied the structure and stability of dispersions containing nanocubes and polymers. Demixing of the cubes and polymers was detected from the apparent static structure factor of mixtures of nanocubes and added polymers, in line with theoretical predictions of colloidal cubes upon addition of non-adsorbing polymers. The obtained experimental phase boundaries are in close agreement with theoretical predictions obtained from free volume theory for mixtures of hard superballs with penetrable hard-sphere depletants. Using static light scattering (SLS) it was possible to determine the phase transition concentrations of mixtures of silica nanocubes and polystyrene in \textit{N,-N-}dimethylformamide containing $40$ mM LiCl. By diluting the phase separated mixture a stable dispersion was recovered, demonstrating that the phase separation is mediated by the presence of non-adsorbing polymers. The results presented here can be a suitable starting point for further experiments on the phase behaviour of nanocube fluids. SLS is a suitable technique to monitor and study the phase behaviour of cubic colloids mixed with non-adsorbing polymers in a common solvent. The SLS setup that was used in this work is unable to quantify the scattering of the solid phase formed. Therefore, further experiments on the sediment by SAXS or SANS are suggested. Additionally, in this work we present an apparent structure factor, since it is impossible to decouple the (weaker) scattering by the polymers from the scattering by the nanocubes. Here, SAXS or SANS also might provide further insight in the structure of the cubes at intermediate polymer concentrations.

\begin{table*}[t]
\caption{Overview of the measured HSN + polystyrene mixtures. }
\label{tab:1}       
\centering
\begin{tabular}{llll}
\hline\noalign{\smallskip}
Experiment 1 & Experiment 2 & Experiment 3 & Experiment 4  \\
(DEP 1) & (DEP 2) & (DEP 3) & (DEP 4)  \\
\noalign{\smallskip}\hline\noalign{\smallskip}
\begin{tabular}{lll} 
no.   &	HSN &  PS  \\
    & (g/L) & (g/L) \\
\hline
1   & 0.35  & 0     \\
2   & 4.63  & 0     \\
3   & 4.51  & 2.46  \\
4   & 4.34  & 5.99  \\
5   & 4.18  & 9.25  \\
6   & 4.05  & 11.9  \\
7   & 3.92  & 14.6  \\
8   & 3.57  & 14.2 \\
9   & 3.27  & 12.9  \\
\end{tabular} & 
\begin{tabular}{lll} 
no.   &	HSN &  PS  \\
    & (g/L) & (g/L) \\
\hline
1   & 1.8   & 0     \\
2   & 10.0  & 0     \\
3   & 9.66  & 3.27  \\
4   & 9.29  & 6.68  \\
5   & 9.00  & 9.37  \\
6   & 8.63  & 12.8  \\
7   & 7.4   & 11.0  \\
\end{tabular} & 
\begin{tabular}{lll}
no.   &	HSN &  PS  \\
    & (g/L) & (g/L) \\
\hline
1   & 1.8  & 0     \\
2   & 24.4  & 0     \\
3   & 23.2  & 2.05  \\
4   & 22.1  & 3.70  \\
5   & 21.0 & 5.43  \\
6   & 20.3  & 6.66  \\
7   & 19.4  & 7.96  \\
8   & 18.7  & 9.20 \\
9   & 18.0  & 10.3  \\
10   & 17.1  & 11.7 \\
11   & 13.6  & 9.31  \\
\end{tabular} & 
\begin{tabular}{lll} 
no.   &	HSN &  PS  \\
    & (g/L) & (g/L) \\
\hline
1   & 1.7   & 0     \\
2   & 36.7  & 0     \\
3   & 34.9  & 4.41  \\
4   & 33.7  & 7.49  \\
5   & 32.6  & 10.4  \\
6   & 31.2  & 13.8  \\
\end{tabular} \\
\noalign{\smallskip}\hline
\end{tabular}
\end{table*}


\section*{Acknowledgements}
\label{sec:16}
Jan Jurre Harsveld van der Veen and Jesse de Boer are thanked for their work on silica cube dispersions mixed with polystyrene. Joeri Opdam is thanked for his contributions \textcolor{black}{to} the theory for superball--depletant mixtures.

\section*{Authors contributions}
RT and APP proposed the research. FD conducted the experiments, and AGG the theoretical calculations. All authors jointly wrote and discussed the manuscript.

\clearpage
\appendix
\section{Scattering of polystyrene solutions}
\begin{figure*}
  \includegraphics[width=0.8\linewidth]{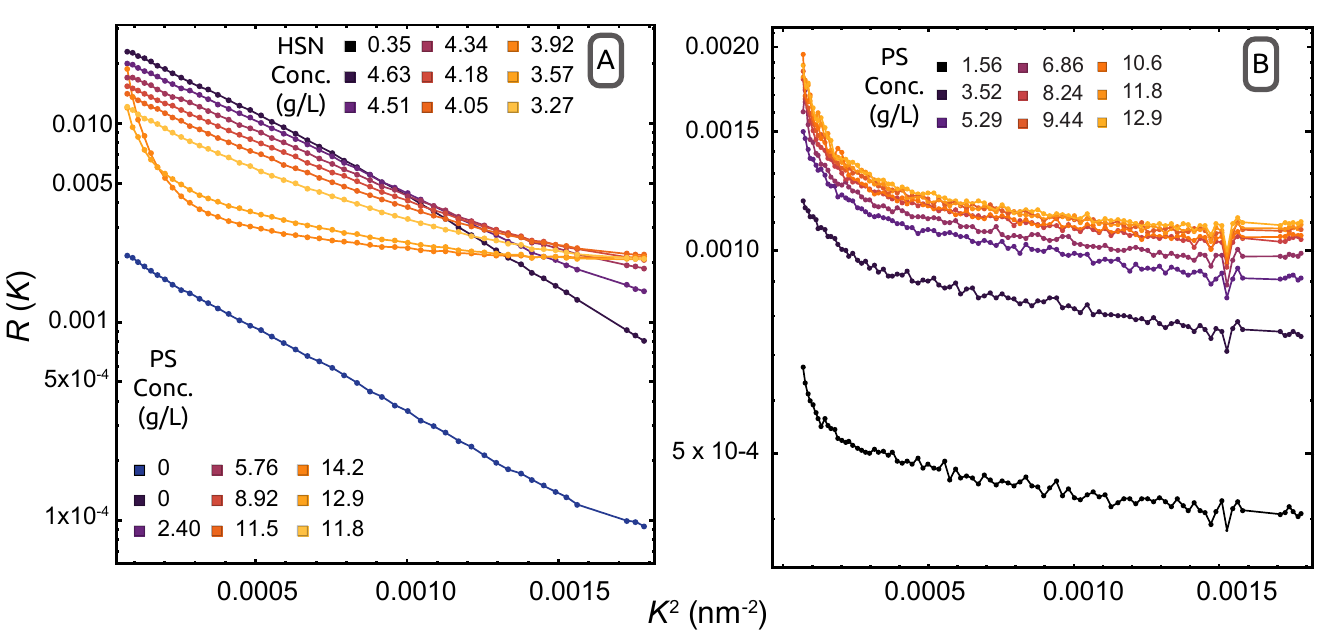}
  \centering
\caption{A: Guinier plot of hollow silica cubes with increasing polystyrene concentration. \textcolor{black}{The HSN and PS concentrations are listed in Table \ref{tab:1}, Experiment 1.} B: Guinier plot of polystyrene with concentrations ranging from 1.56 to 12.9 g/L.}
\label{fig:s1}       
\end{figure*}

In fig. \ref{fig:s1} Guinier plots are plotted for the scattering experiments of the lowest particle concentration and the range of polystyrene concentrations employed in our study. It is evident that the scattering intensity of particles at low wave vectors $K$ is roughly an order of magnitude higher than of the polystyrene solution without cubes. Only at large $K$ values the scattering intensity of the polystyrene attains intensity values comparable to the scattering of the nanocubes.

\bibliographystyle{unsrt}
\bibliography{references}
\end{document}